# Distribution Functions From Breit-Frame Regularisation

Helmut Kröger and Norbert Scheu[a]

[a]Département de Physique, Université Laval, Québec, Québec G1K 7P4, Canada.

We suggest a new Hamiltonian lattice approach, using a regularisation motivated by deep inelastic scattering. We discuss the relation between distribution functions and the $F_1$ structure function. We have tested the method by computing the critical behaviour of the scalar model and find agreement with scaling behaviour and with results by Lüscher and Weisz.

The computation of *quark distribution functions, structure functions* and of the *hadronic mass spectrum* directly from QCD requires nonperturbative methods. The most important non perturbative approach to compute these quantities is lattice gauge theory. Lattice gauge theory can be formulated in both Euclidean space or Minkowski space. Here we shall explore the usefulness of the Hamiltonian or transfer-matrix approach in Minkowski space as an alternative to the commonly used statistical methods in Euclidean space[1]. From a *numerical* point of view, however, it is extremely difficult to actually diagonalise a Hamiltonian ( or transfer matrix) even on very small lattices with the currently available numerical techniques. *Euclidean lattice gauge theory* is up to now the only nonperturbative method which is capable of numerically describing the low-energy part of the observed hadronic mass spectrum directly from $QCD$ (Glueball masses, however, have already been computed in the Hamiltonian approach[1]). But unfortunately, the computation of distribution functions or structure functions is very difficult in Euclidean space. So far, the first moments of *valence quark* structure functions can be measured. Important work in this direction has been done by Schierholz and co-workers[2]. The computation of the first moments of the *sea quark* contribution to the structure functions is feasible in principle but in practice very complicated and has not been done yet. A computation of quark distribution functions beyond the first moments is still beyond reach of the currently available methods. The computation of distribution functions (not only the first moments thereof) would be relatively simple if it were possible to obtain the hadronic wave functions from a Minkowski space calculation. Two other problems of the Euclidean approach

– finite density thermodynamics and
– excited states of mass spectra

are expected to be simpler in the *Minkowskian* approach, if it were possible to diagonalise the transfer matrix of the field theory under investigation. A further advantage of this approach is that it is closer to the *intuitive picture* of a Hadron as a many body bound state of quarks and gluons with wave functions describing the probability amplitude of finding a certain configuration of quarks and gluons inside the proton. It is, however, this very advantage which causes the numerical problems because a hadron consists of a large avarage number of constituents which diverges in the thermodynamic limit. A many particle system is very difficult to treat numerically and even more so a system with fluctuating (non conserved) constituent number. To render things worse, *mean field* many body techniques (e.g. Hartree-Fock) are *not* viable for a relativistic field theory in fewer than $4 + 1$ space-time dimensions.

So how could the $QCD$ Hamiltonian be diagonalised with reasonable numerical effort? Our idea is to consider experiment as a guide-line. More specificly we consider deep inelastic scattering experiments where the composite nature of hadrons becomes visible. Our aim is to describe only properties of the constituents in the the hadron than we are actually able to measure. This has three immediate consequences. *Firstly*, the observed quarks depend on the experimental resolution, according to the renormalisation group improved parton model. They are effective particles, not elementary ones. Quarks observed at low resolution consist of quarks and gluons at a higher resolution scale. Hence the lattice spacing has to be chosen to be of the same order of magnitude as the experimental resolution. *Secondly*, the resolution of the experiment (= the wavelength $\frac{\lambda}{2\pi} = 1/|\vec{q}|$ of the exchanged virtual photon with four-momentum $q$ ) is *not* Lorentz invariant and so are the effective quarks. In the Breit-frame (where the hadron four-momentum takes on the form $P = (E, 0.0, P_3)$ and where the momentum of the virtual photon $q = (0,0,0,-Q)$) the inverse resolution coincides with the relativistic invariant $Q := \sqrt{-q^2}$. In this frame, the Bjorken scaling variable $x := \frac{Q^2}{2P \cdot q}$ can be interpreted as the fraction $x = \frac{p_3}{P_3}$ of the momentum $\vec{p}$ of the struck parton inside the total proton momentum $P_3$. Therefore we chose this frame as the lattice rest-frame. *Finally* and most importantly the observable parton momenta necessarily lie in the interval $0 < p_3 < P_3$ if the hadron is stable. Hence, for *every* finite lattice regularisation the number of observable states in the Fock space is finite. The mathematical reason for this is that only a finite number of particles with discrete momenta can share a given total momentum. The *infinite momentum frame* shares a similar feature. There is, however, a major advantage of using a high momentum frame which is *not* the infinite momentum frame: We have a finite $space-like$ lattice spacing at our disposition which is not possible in the front-form nor is it possible with infinite hadron momentum(!) This is an important feature because without a finite lattice spacing, no dimensionless correlation length $\xi = \frac{1}{m_R a}$ can be defined from the hadron mass $m_R$. In the continuum limit $\xi \to \infty$, however, we end up with an infinite hadron momentum since the hadron velocity $v = P_3/E = P_3 a/\sqrt{1/\xi^2 + P_3^2 a^2} \to 1$ approaches the velocity of light. The same thing holds for the *Bjorken limit* $Q \to \infty, x = const$, since in the Breit frame the parton momenta $p_3 = Q/2 \to \infty$ as well as the proton momentum $P_3 = Q/2/x \to \infty$, diverge in this limit.

---

[1]Both approaches should be equivalent if the measured correlation length is much greater than the lattice spacing For a small correlation length, a "renormalisation of the speed of light" can be necessary.

As a first test of the Breit method, we have applied it to the $\phi^4$ model in four space-time dimensions. The Lagrangian density of this model is
$$L = \frac{1}{2}\partial_\mu \phi \partial^\mu \phi - \frac{1}{2}m^2\phi^2 - \frac{g}{4!}\phi^4.$$
In Euclidean lattice calculations, the mass $m$ and the coupling constant $g$ are usually replaced by the hopping parameter $\kappa$ and the lattice coupling constant $\lambda$ as follows $m^2 = \frac{1-2\lambda}{\kappa} - 8, g = \frac{6\lambda}{\kappa^2}$. This model shows two phases: the asymmetric phase and the symmetric one. We considered the latter. As one approaches the critical line in the $\kappa, \lambda$ parameter space, the correlation length $\xi = \frac{1}{m_R a}$ diverges, where $m_R$ is the mass of the ground state. Close to the critical line, the ground state masses should obey the following universal scaling law:
$$1/\xi = m_R a \sim C\tau^{\frac{1}{2}}|\log \tau|^{-1/6}$$
where $\tau = \frac{\kappa - \kappa_c}{\kappa_c}$ and $\kappa_c$ is the critical couplong where the correlation length diverges. Fig.[1] compares our results to semianalytic calculations[3] showing that the asymptotic scaling law is nicely reproduced. In Fig.[2] the mass ratios of the spectrum are displayed. We observe scaling close to the critical point.

One comment on the scalar $\phi^4$ theory is in order: If we had used the zero-momentum sector where the bound state is at rest, the problem would have been a complicated many-body problem. The 'parton cloud' would have appeared[4]. In a fast-moving frame, however, the one-particle sector dominates the bound-state.

We have designed our method in view of $QCD$. Therefore, it is useful to calculate the structure functions in our approximation. A straightforward calculation using the impulse approximation yields the following expression

$$F_1(x,Q) = \sum_{i,s} e_i^2/2 \{f_s^{(i)}(x, 1, \frac{2x\mu}{Q}) + \bar{f}_s^{(i)}(x, 1, \frac{2x\mu}{Q})\}$$

where $f_s^{(i)}(p_3, P_3, \mu) := \int d^2 p_\perp B(\vec{p}) w_s^{(i)}(\vec{p}, P_3, \mu)$ and where $w_s^{(i)}(\vec{p}, P_3, \mu) := \frac{<PS|b_{i,s}^{\dagger}(\vec{p})b_{i,s}(\vec{p})|PS>}{<PS|PS>}$ is the probability of finding a quark with helicity $s$ and momentum $\vec{p}$ inside a hadron with four-momentum $P$ and Pauli-Lubanski spin vector $S$. In analogy, $\bar{w}$ and $\bar{f}$ refers to an antiquark. The quantity $B$ is defined as $B(\vec{p}) := \frac{E}{2} \frac{\vec{p}^2}{p_3 \sqrt{m^2 + \vec{p}^2}}$. If we neglect parton masses and transverse momenta in the Bjorken limit we end up with the well known interpretation of the $F_1$ structure function as a linear combination of quark distribution functions $q_{s,i}(x,Q) = f_s^{(i)}(x, 1, 2x\mu/Q) = P_3 f_s^{(i)}(p_3, P_3, \mu)$. The same holds for $F_2$ and $g_1$.

In conclusion, the diagonalisation of a relativistic Hamiltonian is much simpler in a large momentum sector than in the zero momentum sector. We have tested this for the case of the $\phi^4$-model where we obtained asymptotic scaling in good agreement with the semianalytical results of Lüscher and Weisz. In deep inelastic scattering this corresponds to taking into account partons with experimentally observable momenta only.

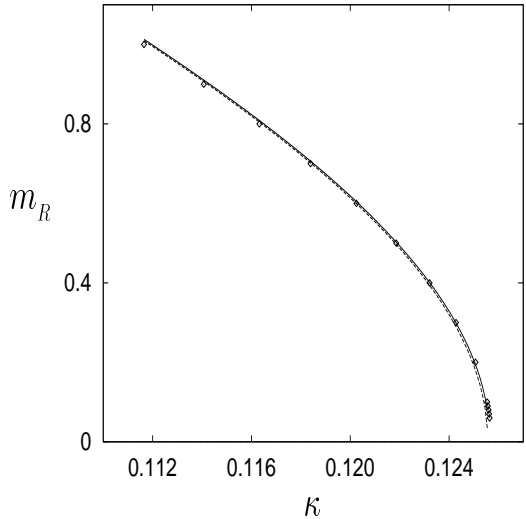

**Fig.1** The ground state mass $m_R$ in lattice units ($a \equiv 1$) versus $\kappa$ for $\lambda = 0.00345739$ ($\tilde{\lambda} = 0.01$ in Ref.[3]). The dots correspond to results of Ref.[3]. Our results correspond to a $7^3$ lattice (dashed line) and to a $9^3$ lattice (solid line).

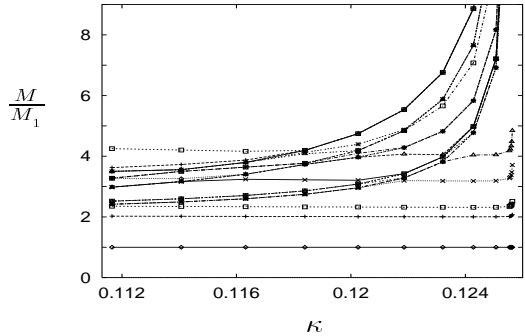

**Fig.2** The lowest lying mass spectrum versus $\kappa$. The ground state mass is set to one. $\lambda$ as in Fig.[1].